\documentclass[aps,preprint]{revtex4}%
\usepackage{amsfonts}
\usepackage{amsmath}
\usepackage{amssymb}
\usepackage{subfigure}
\usepackage{graphicx}%
\begin{document}
\preprint{CTP-SCU/2016011}
\title{Closed String Tachyon Driving $f(R)$ Cosmology}
\author{Peng Wang$^{a}$}
\email{pengw@scu.edu.cn}
\author{Houwen Wu$^{a,b}$}
\email{wu_houwen@g.harvard.edu}
\author{Haitang Yang$^{a}$}
\email{hyanga@scu.edu.cn}
\affiliation{$^{a}$Center for Theoretical Physics, College of Physical Science and
Technology, Sichuan University, Chengdu, 610064, PR China}
\affiliation{$^{b}$Center for the Fundamental Laws of Nature, Harvard University,
Cambridge, MA, 02138, USA}

\begin{abstract}
To study quantum effects on the bulk tachyon dynamics, we replace $R$ with
$f(R)$ in the low-energy effective action that couples gravity, the dilaton,
and the bulk closed string tachyon of bosonic closed string theory and study
properties of their classical solutions. The $\alpha^{\prime}$ corrections of
the graviton-dilaton-tachyon system are implemented in the $f(R)$. We obtain
the tachyon-induced rolling solutions and show that the string metric does not
need to remain fixed in some cases. The singular behavior of more classical
solutions are investigated and found to be modified by quantum effects. In
particular, there could exist some classical solutions, in which the tachyon
field rolls down from a maximum of the tachyon potential while the dilaton
expectation value is always bounded from above during the rolling process.

\end{abstract}
\keywords{}\maketitle
\tableofcontents



\section{Introduction}

In the standard cosmological model, the Friedmann equations are derived from
Hilbert action, coupling the metric with ad hoc matter sources. In string
theory, a low energy effective action arises from the requirement of quantum
conformal invariance. The tree level of this action couples the metric, the
dilaton and the axion. This action, implying the dynamics of gravitational
field and other background fields, has been used in many areas. One of the
most important applications is to develop a string cosmology
\cite{Gasperini:2002bn, Battefeld:2005av}.

There exist two ways to improve this string effective action
\cite{Green:1987sp, Polchinski:1998rq}. The first one is $\alpha^{\prime}%
$-controlled expansion, which includes higher derivatives of the metric and
background fields. Another one is string coupling $g_{s}$-controlled expansion
(higher-genus expansion) and which reflects the higher loop string
interactions. The $\alpha^{\prime}$-controlled expansion, which becomes
significant in Planck and high curvature region, had been discussed in
\cite{high curvature WDW 1, high curvature WDW 2}. When curvature and energy
scale grow up, the usual simplest perturbative expansion of the string
effective action becomes no-go theorem. In this condition, the higher order
expansions should be taken into account. A good review of string cosmology is
given by \cite{Gasperini:2007zz} and references therein.

The instabilities associated with open string tachyon is well understood
presently. The stable vacuum is the vacuum of closed string without open
string excitations and D-branes. The open string tachyon rolls down from the
perturbative unstable vacuum to an excited state of closed string carrying the
energy of the D-brane. On the other hand, the instabilities associated with
closed string tachyon is a much more difficult problem since the action is
non-polynomial. Some progresses have been achieved in the calculation of the
effective potential \cite{Yang:2005rx, Moeller:2006cv, Moeller:2006cw,
Moeller:2007mu} by truncating the polynomial to quintic order. Those
calculations indicate that there do exist a local minimum and several saddle
points in the closed string tachyon and dilaton effective potential. The
conjecture that the action has to vanish at the closed string field theory
(CSFT) vacuum, made in \cite{Yang:2005rx, Yang:2005rw}, is also implied. At
the closed string vacuum, parallel to the conclusions in open string theory,
one is led to believe that spacetime itself ceases to exist. To make these
conjectures more reliable, higher order calculations or more desirable
analytical methods are necessary.

For these reasons, studying the string effective action that couples the
metric, the dilaton and the tachyon is of importance. One can expect that this
action will reveal some features of string theory and related problems. Some
progresses have been made in recent developments. The author of
\cite{Suyama:2006wx} discussed the solutions of the effective action with
tachyon and B-field and analyzed the solutions in $AdS_{3}$ background. In
\cite{Brandenberger:2007xu}, Brandenberger \emph{et.al.} found a nonsingular
and static tachyon condensation by discussing an effective theory with a
non-vanishing dilaton potential. The quantum effects of
graviton-dilaton-tachyon system is investigated in \cite{Suyama:2008hk} where
it shows that the singular behavior of classical solutions should be modified
by quantum effects. The author of \cite{Headrick:2008sa} uses some constraints
to fix the form of action without computing the beta functions. Some other
progresses of graviton-dilaton-tachyon system refer to
\cite{Alexandre:2008sr,Hikida:2005ec,Singh:2005hb, Suyama:2005wd,
Genenberg:2006de,Bergman:2006pd,Razamat:2007ky,
Swanson:2008dt,Aref'eva:2008gj,EscamillaRivera:2011di}.

The aim of this paper is an extension of \cite{Yang:2005rx,Yang:2005rw} to
study the $\alpha^{\prime}$ corrections of the graviton-dilaton-tachyon
system. The $\alpha^{\prime}$ correction should also satisfy the quantum
conformal invariance and then brings higher derivative terms of gravitational
field as well as other background fields. To preserve the covariance and gauge
invariance, the same expansion order of $\alpha^{\prime}$ could lead to
different effective actions \cite{Gasperini:2007zz,RA}. To simplify the story,
we only implement corrections to the graviton by replacing $R$ with $f(R)$.
This could be accomplished by integrating out other massive fields in the action.

In \cite{Yang:2005rw}, the rolling process triggered by the tachyon was
investigated for tachyonic potentials. It was found that during the rolling
process, the string metric did not evolve while the dilaton rolled to strong
coupling. In the framework of $f(R)$ gravity, our analysis shows some
interesting results. In section \ref{Sec:TRS}, we find for tachyon-induced
rolling solutions that time evolution of the string metric could be nontrivial
in some cases. In section \ref{Sec:CS}, assuming the Hubble parameter of the
string metric is constant, we obtain a set of solutions, in which the tachyon
field rolls down from the top of some tachyon potential while the string
coupling is always finite during the rolling process.

This paper is organized as follows. In section \ref{Sec:CSF}, we derive the
equations of motion of graviton, dilaton, and tachyon field in $f\left(
R\right)  $ theories. In section \ref{Sec:TRS}, we define tachyon-induced
rolling solutions and solve the equations of motion for them. More classical
solutions are discussed in section \ref{Sec:CS}. The section \ref{Sec:CON} is
our conclusion.

\section{Coupled System of Fields}

\label{Sec:CSF}

The low energy effective action for the metric, the dilaton, and the tachyon
is given by%

\begin{equation}
S=\frac{1}{2\kappa^{2}}\int d^{d+1}x\sqrt{-g}e^{-2\Phi}\left[  f\left(
R\right)  +4\left(  \nabla\Phi\right)  ^{2}-\left(  \nabla T\right)
^{2}-2V\left(  T\right)  \right]  , \label{eq:action}%
\end{equation}

\noindent where $g_{\mu\nu}$ is string metric, $\Phi$ is dilaton, $T$ is
tachyon with potential $V\left(  T\right)  $, and $f\left(  R\right)  $ is the
generalized gravity. The number of spatial dimensions is $d$. Since here we
focus on the tachyon driving solutions, the dilaton potential is set to zero
for simplicity. Varying the action $\left(  \ref{eq:action}\right)  $ with
respect to $g_{\mu\nu}$, $\Phi$, and $T$, we find that the equations of motion
for graviton, dilaton and tachyon are%
\begin{gather}
F\left(  R\right)  R_{\mu\nu}-\nabla_{\mu}T\nabla_{\nu}T+4g_{\mu\nu}\left[
F\left(  R\right)  -1\right]  \left(  \nabla\Phi\right)  ^{2}-4g_{\mu\nu
}\nabla^{\alpha}\Phi\nabla_{\alpha}F\left(  R\right) \nonumber\\
-2g_{\mu\nu}\left[  F\left(  R\right)  -1\right]  \nabla^{2}\Phi+g_{\mu\nu
}\nabla^{2}F\left(  R\right)  +2F\left(  R\right)  \nabla_{\nu}\nabla_{\mu
}\Phi\nonumber\\
-\nabla_{\nu}\nabla_{\mu}F\left(  R\right)  -4\left[  F\left(  R\right)
-1\right]  \nabla_{\nu}\Phi\nabla_{\mu}\Phi+2\nabla_{\mu}\Phi\nabla_{\nu
}F\left(  R\right)  +2\nabla_{\nu}\Phi\nabla_{\mu}F\left(  R\right)
=0,\label{eq:equation of motion 1}\\
\frac{1}{2}\left[  F\left(  R\right)  R-f\left(  R\right)  \right]  +2\left\{
d\left[  F\left(  R\right)  -1\right]  +1\right\}  \left(  \nabla\Phi\right)
^{2}-2d\nabla^{\alpha}\Phi\nabla_{\alpha}F\left(  R\right) \nonumber\\
-\left\{  d\left[  F\left(  R\right)  -1\right]  +1\right\}  \nabla^{2}%
\Phi+\frac{1}{2}d\nabla^{2}F\left(  R\right)  +V\left(  T\right)
=0,\label{eq:equation of motion 2}\\
\nabla^{2}T-2\nabla^{\mu}\Phi\nabla_{\mu}T-V^{\prime}\left(  T\right)  =0,
\label{eq:equation of motion 3}%
\end{gather}
where $F(R)\equiv df(R)/dR$. As in \cite{Yang:2005rw}, we make the following
ansatz%
\begin{align}
ds^{2}  &  =-dt^{2}+a\left(  t\right)  ^{2}\delta_{ij}dx^{i}dx^{j},\nonumber\\
\Phi &  =\Phi\left(  t\right)  ,\label{eq:ansatz}\\
T  &  =T\left(  t\right)  .\nonumber
\end{align}
For the string metric in eqn. $\left(  \ref{eq:ansatz}\right)  $, the Ricci
scalar is%
\begin{equation}
R=2d\dot{H}+d\left(  d+1\right)  H^{2},
\end{equation}
where $H=\frac{\dot{a}}{a}$. In this case, the $00$ and $ij$ components of
gravitational equations become%

\begin{gather}
F\left(  R\right)  \left(  d\dot{H}+dH^{2}\right)  +\dot{T}^{2}-2\ddot{\Phi
}+\left\{  2\left[  F\left(  R\right)  -1\right]  \dot{\Phi}-\dot{F}\left(
R\right)  \right\}  dH=0,\label{eq: 00 component}\\
F\left(  R\right)  \left(  \dot{H}+dH^{2}\right)  -4\left[  F\left(  R\right)
-1\right]  \dot{\Phi}^{2}+4\dot{\Phi}\dot{F}\left(  R\right)  +2\left[
F\left(  R\right)  -1\right]  \ddot{\Phi}\nonumber\\
+2\left[  F\left(  R\right)  \left(  d-1\right)  -d\right]  H\dot{\Phi}%
-\ddot{F}\left(  R\right)  -\left(  d-1\right)  H\dot{F}\left(  R\right)  =0,
\label{eq:ij component}%
\end{gather}
which can be rearranged into two equivalent equations%
\begin{gather}
\frac{1}{2}F\left(  R\right)  \left(  d-1\right)  \dot{H}+\frac{1}{2}\dot
{T}^{2}+2\left[  F\left(  R\right)  -1\right]  \dot{\Phi}^{2}-2\dot{\Phi}%
\dot{F}\left(  R\right) \nonumber\\
-F\left(  R\right)  \ddot{\Phi}+F\left(  R\right)  H\dot{\Phi}+\frac{1}%
{2}\ddot{F}\left(  R\right)  -\frac{1}{2}H\dot{F}\left(  R\right)
=0,\label{eq:mix 1}\\
\frac{1}{2}F\left(  R\right)  \left(  d-1\right)  dH^{2}-2d\left[  F\left(
R\right)  -1\right]  \dot{\Phi}^{2}+2d\dot{\Phi}\dot{F}\left(  R\right)
+\left\{  d\left[  F\left(  R\right)  -1\right]  +1\right\}  \ddot{\Phi
}\nonumber\\
+\left\{  \left[  d\left(  F\left(  R\right)  -1\right)  -2F\left(  R\right)
+1\right]  \dot{\Phi}-\frac{1}{2}\left(  d-2\right)  \dot{F}\left(  R\right)
\right\}  dH-\frac{1}{2}d\ddot{F}\left(  R\right)  -\frac{1}{2}\dot{T}^{2}=0.
\label{eq:mix 2}%
\end{gather}
The equations of motion for the dilaton and the tachyon are%
\begin{gather}
\frac{1}{2}\left[  F\left(  R\right)  R-f\left(  R\right)  \right]  -2\left[
d\left(  F\left(  R\right)  -1\right)  +1\right]  \dot{\Phi}^{2}+2d\dot{\Phi
}\dot{F}\left(  R\right) \nonumber\\
+\left\{  d\left[  F\left(  R\right)  -1\right]  +1\right\}  \left(
\ddot{\Phi}+dH\dot{\Phi}\right)  -\frac{1}{2}d\left[  \ddot{F}\left(
R\right)  +dH\dot{F}\left(  R\right)  \right]  +V\left(  T\right)
=0,\label{eq: dilatoneom}\\
\ddot{T}+\left(  dH-2\dot{\Phi}\right)  \dot{T}+V^{\prime}\left(  T\right)
=0. \label{eq:tachyoneom}%
\end{gather}
Note that there are four differential equations for three unknown dynamical
variables: $a\left(  t\right)  $, $\Phi\left(  t\right)  $, and $T\left(
t\right)  $. Therefore, one of the four equations should be redundant. In
fact, it shows in the appendix that eqns. $(\ref{eq: 00 component})$,
$(\ref{eq:ij component})$, and $(\ref{eq:tachyoneom})$ could guarantee that
eqn. $(\ref{eq: dilatoneom})$ holds whenever $\dot{T}\neq0$.

\section{Tachyon-Driven Rolling Solutions}

\label{Sec:TRS}

In this section, we consider a general class of potentials $V(T)$ for a
tachyon $T$ that has a local maximum at $T=0$, which can be written as%

\begin{equation}
V\left(  T\right)  =V(0)-\frac{1}{2}m^{2}T^{2}+\mathcal{O}\left(
T^{3}\right)  .
\end{equation}
In \cite{Yang:2005rw} where $f\left(  R\right)  =R$, the rolling solutions
driven by the tachyon were discussed. The ansatzes for $T\left(  t\right)  $
and $\Phi\left(  t\right)  $ were assumed to be%
\begin{align}
T\left(  t\right)   &  =e^{mt}+\underset{n\geq2}{\sum}t_{n}e^{nmt},\nonumber\\
\Phi\left(  t\right)   &  =\underset{n\geq2}{\sum}\phi_{n}e^{nmt}%
,\label{eq:TPhiHansatz}\\
H\left(  t\right)   &  =\sum_{n\geq2}h_{n}e^{n\gamma t},\nonumber
\end{align}
where $T\rightarrow0$ for $t\rightarrow-\infty$, and $\Phi\left(  t\right)  $
has exponentials subleading to $e^{mt}$ since the tachyon drives the rolling
in the very early time. For the case with $f\left(  R\right)  =R$, eqn.
$\left(  \ref{eq:mix 2}\right)  $ becomes%
\begin{equation}
\frac{1}{2}\left(  d-1\right)  dH^{2}=\frac{1}{2}\dot{T}^{2}-\ddot{\Phi
}+dH\dot{\Phi}, \label{eq:HVanish}%
\end{equation}
which gives that $H\rightarrow0$ when $t\rightarrow-\infty$, which is
consistent with the ansatz for $H\left(  t\right)  $ proposed in eqn. $\left(
\ref{eq:TPhiHansatz}\right)  $. In fact, it showed in \cite{Yang:2005rw} that
$H\left(  t\right)  $ vanished identically for the tachyon-driven rolling solutions.

For the case with a general form of $f\left(  R\right)  $, we build an
analogous tachyon-driven rolling solution:%
\begin{align}
T\left(  t\right)   &  =e^{\gamma t}+\underset{n\geq2}{\sum}t_{n}e^{n\gamma
t},\nonumber\\
\Phi\left(  t\right)   &  =\underset{n\geq2}{\sum}\phi_{n}e^{n\gamma t},
\label{eq:T-phi ansatz}%
\end{align}
where $\gamma$ is a positive real number, and the first term in $T\left(
t\right)  $ is the solution to the linearized tachyon equation of motion. The
arbitrary constant multiplying the first term in $T\left(  t\right)  $ can be
absorbed, as we did, by a redefinition of time.\ Unlike in the case with
$f\left(  R\right)  =R$, eqn. $\left(  \ref{eq:mix 2}\right)  $ fails to
require that $H\rightarrow0$ when $t\rightarrow-\infty$. There might exist
possible solutions with $H\left(  t\right)  $ which does not go to zero when
$t\rightarrow-\infty$. Thus, one might consider a more general ansatz for
$H\left(  t\right)  $:
\begin{equation}
H\left(  t\right)  =H_{0}+\sum_{n\geq1}h_{n}e^{n\gamma t}. \label{eq:H ansatz}%
\end{equation}
If the rolling process is solely trigger by the tachyon filed, one needs that
$H_{0}=h_{1}=0$. As we show below, for such ansatz nontrivial $H\left(
t\right)  $ could exist in some cases. However to explore more possibilities,
we now do not require $H_{0}=h_{1}=$ $0$. Plugging $T\left(  t\right)  $ and
$\Phi\left(  t\right)  $ from eqns. $\left(  \ref{eq:T-phi ansatz}\right)  $
into the tachyon equation $\left(  \ref{eq:tachyoneom}\right)  $, we find%
\begin{equation}
\gamma^{2}+dH_{0}\gamma-m^{2}=0\text{,}%
\end{equation}
which gives $\gamma=m\left(  \sqrt{1+\frac{d^{2}H_{0}^{2}}{4m^{2}}}%
-\frac{dH_{0}}{2m}\right)  $. The corresponding Ricci scalar is
\begin{equation}
R=R_{0}+\sum_{n\geq1}r_{n}e^{n\gamma t}, \label{eq: R ansatz}%
\end{equation}
where $R_{0}=d\left(  d+1\right)  H_{0}^{2}$ and $r_{n}$ depends on $h_{m\leq
n}$. We assume $f\left(  R\right)  $ is analytic along the real axis except
certain poles.

In the rest of this section, we will use eqns. $\left(  \ref{eq: 00 component}%
\right)  $, $\left(  \ref{eq:ij component}\right)  $ and $(\ref{eq:tachyoneom}%
)$ to solve for $R_{0}$, $h_{n}$, $\phi_{n}$ and $t_{n}$. And eqn.
$(\ref{eq: dilatoneom})$ could be used to determine the value of $V\left(
0\right)  $. Note that in \cite{Yang:2005rw} where $f\left(  R\right)  =R$, it
showed that $V\left(  0\right)  =0$, $R_{0}=0$ and $h_{n}=0$.

\subsection{Analytic at $R_{0}$}

\label{subSec:R0 analytic}

Now $f\left(  R\right)  $ is assumed to be analytic at $R_{0}$. Thus, we can
expand $f\left(  R\right)  $ at $R=R_{0}$:
\begin{equation}
f\left(  R\right)  =f\left(  R_{0}\right)  +\underset{l\geq1}{\sum}\alpha
_{l}\left(  R-R_{0}\right)  ^{l}.
\end{equation}
Eqn. $\left(  \ref{eq: R ansatz}\right)  $ gives $F\left(  R\right)  =F\left(
R_{0}\right)  +\mathcal{O}\left(  e^{\gamma t}\right)  $. Thus, the leading
terms of the $00$ and $ij$ components of gravitational equations $\left(
\ref{eq: 00 component}\right)  $ and $\left(  \ref{eq:ij component}\right)  $
both become $F\left(  R_{0}\right)  dH_{0}^{2}=0$. Given $R_{0}=d\left(
d+1\right)  H_{0}^{2}$, one finds that either $F\left(  R_{0}\right)  =0$ or
$R_{0}=0.$ The tachyon equation $\left(  \ref{eq:tachyoneom}\right)  $ is
trivial at the leading order. The leading order of the dilation equation
$\left(  \ref{eq: dilatoneom}\right)  $ gives%
\begin{equation}
V\left(  0\right)  =-\frac{f\left(  R_{0}\right)  }{2}, \label{eq:V0}%
\end{equation}
where we use $F\left(  R_{0}\right)  R_{0}=0$. Note that eqn. $\left(
\ref{eq:V0}\right)  $ must be satisfied to admit the tachyon rolling solution.
For example, if $f\left(  R\right)  =R-\Lambda$ where $\Lambda$ is the
cosmological constant and $R_{0}=0$, one has $V\left(  0\right)
=\frac{\Lambda}{2}$. In what follows, we will solve for $h_{n}$, $\phi_{n}$
and $t_{n}$ in the two cases with $R_{0}=0$ and $F\left(  R_{0}\right)  =0$.

\subsubsection{$R_{0}=0$}

If $f\left(  R\right)  $ is analytic at $R=0$, $F\left(  R\right)
=1+\underset{l\geq2}{\sum}l\alpha_{l}R^{l-1}$. Plugging eqns. $\left(
\ref{eq:T-phi ansatz}\right)  $ and $\left(  \ref{eq:H ansatz}\right)  $ with
$H_{0}=0$ into the equations of motion, one could determine the coefficients
$t_{n},\phi_{n}$ and $h_{n}$ order by order. Note that $\gamma=m$ for
$H_{0}=0$.

At $\mathcal{O}\left(  e^{mt}\right)  ,$ eqns. $\left(  \ref{eq: 00 component}%
\right)  $ and $\left(  \ref{eq:ij component}\right)  $ give%
\begin{align}
dmh_{1}e^{mt}  &  =0,\nonumber\\
\left(  1-4dm^{2}\alpha_{2}\right)  mh_{1}e^{mt}  &  =0,
\end{align}
which leads to $h_{1}=0$. Tachyon equation is also trivial at this order. At
$\mathcal{O}\left(  e^{2mt}\right)  ,$eqns. $\left(  \ref{eq: 00 component}%
\right)  $ and $\left(  \ref{eq:ij component}\right)  $ give%
\begin{align}
\left(  m^{2}+2dmh_{2}-8m^{2}\phi_{2}\right)  e^{2mt}  &  =0,\nonumber\\
2m\left(  1-16dm^{2}\alpha_{2}\right)  h_{2}e^{2mt}  &  =0,
\end{align}
which yield%
\begin{align}
\phi_{2}  &  =\frac{1}{8}+\frac{dh_{2}}{4m},\nonumber\\
h_{2}  &  =0\text{ or }\alpha_{2}=\frac{1}{16dm^{2}}\text{.}%
\end{align}
Note that we have $h_{2}=0$ if $\alpha_{2}\neq\frac{1}{16dm^{2}}$. Now we
prove by induction that $H(t)$ vanishes identically if $\alpha_{2}\neq\frac
{1}{4dN^{2}m^{2}}$ for $N=2,3\cdots$. First assume that $h_{2}=h_{3}%
=\cdots=h_{N}=0$. Since $\dot{\Phi}\left(  t\right)  \sim e^{2mt}$, eqn.
$\left(  \ref{eq:ij component}\right)  $ gives
\begin{equation}
\left(  N+1\right)  \left[  1-4d\alpha_{2}\left(  N+1\right)  ^{2}%
m^{2}\right]  mh_{N+1}e^{\left(  N+1\right)  mt}+\mathcal{O}\left(  e^{\left(
N+2\right)  mt}\right)  =0,
\end{equation}
which yields $\left[  1-4d\alpha_{2}\left(  N+1\right)  ^{2}m^{2}\right]
h_{N+1}=0$. Therefore, if $\alpha_{2}\neq\frac{1}{4dN^{2}m^{2}}$ for
$N=2,3\cdots$, $h_{N+1}=0$.

On the other hand, if $\alpha_{2}=\frac{1}{4dN^{2}m^{2}}$ for some positive
integer $N>1$, $h_{N}$ could be nonzero. Therefore, there might be solutions
with nonzero $H(t)$ in such cases. Consider an example with
\[
f\left(  R\right)  =R+\frac{R^{2}}{16dm^{2}}+\sum\limits_{l\geq3}\alpha
_{l}R^{l}\text{ and }V\left(  T\right)  =-\frac{1}{2}m^{2}T^{2},
\]
we find solutions up to $\mathcal{O}\left(  e^{4mt}\right)  $
\begin{align}
T\left(  t\right)   &  =e^{mt}+\frac{e^{3mt}}{16}+\mathcal{O}\left(
e^{4mt}\right)  ,\nonumber\\
\Phi\left(  t\right)   &  =\left(  \frac{1}{8}+\frac{dh}{4m}\right)
e^{2mt}+\mathcal{O}\left(  e^{4mt}\right)  ,\\
H\left(  t\right)   &  =he^{2mt}+\mathcal{O}\left(  e^{4mt}\right)  ,\nonumber
\end{align}
where $h$ is a free parameter. As far as the tachyon rolling ansatz is
concerned, the parameter $h$ could be arbitrary.

\subsubsection{$F\left(  R_{0}\right)  =0$}

Suppose $F\left(  R\right)  $ is analytic at $R_{0}$ where $F\left(
R_{0}\right)  =0$. Then, one can expand $F\left(  R\right)  $ at $R_{0}$:%
\begin{equation}
F\left(  R\right)  =\underset{l\geq1}{\sum}\beta_{l}\left(  R-R_{0}\right)
^{l},
\end{equation}
where we assume $\beta_{1}\neq0$ for simplicity. Plugging eqns. $\left(
\ref{eq:T-phi ansatz}\right)  $ and $\left(  \ref{eq:H ansatz}\right)  $ into
eqns. $\left(  \ref{eq: 00 component}\right)  $, $\left(
\ref{eq:ij component}\right)  $, and $(\ref{eq:tachyoneom})$ gives the
recurrence relations for $h_{n},\phi_{n},$and $t_{n}$ with necessary initial
values. In fact, after putting the ansatzes into eqns. $\left(
\ref{eq: 00 component}\right)  $, $\left(  \ref{eq:ij component}\right)  $,
and $(\ref{eq:tachyoneom})$, one has at $\mathcal{O}\left(  e^{n\gamma
t}\right)  $ that
\begin{gather}
2n\gamma\left[  \left(  n\gamma+dH_{0}\right)  \phi_{n}-d^{2}H_{0}^{2}%
\beta_{1}h_{n}\right]  =G_{n}\left(  h_{i<n},\phi_{i<n},t_{i<n}\right)
,\nonumber\\
2n\gamma\left(  n\gamma+dH_{0}\right)  \phi_{n}+2n\gamma d\left[  \left(
n^{2}\gamma^{2}-H_{0}^{2}\right)  +\left(  d-1\right)  H_{0}n\gamma\right]
\beta_{1}h_{n}=F_{n}\left(  h_{i<n},\phi_{i<n},t_{i<n}\right)  ,\\
\gamma\left[  \left(  n^{2}-1\right)  \gamma+dH_{0}\left(  n-1\right)
\right]  t_{n}=H_{n}\left(  h_{i<n},\phi_{i<n},t_{i<n}\right)  ,\nonumber
\end{gather}
where $G_{n}$, $F_{n}$ and $H_{n}$ are functions of only $h_{i<n}$,
$\phi_{i<n}$, and $t_{i<n}$. Solving the above equations for $h_{n}$,
$\phi_{n}$, and $t_{n}$, we find that the recurrence relations for $h_{n}%
,\phi_{n},$and $t_{n}$ for $n\geq2$ are
\begin{align}
\phi_{n}  &  =\frac{G_{n}\left(  h_{i<n},\phi_{i<n},t_{i<n}\right)  }%
{2n\gamma\left(  n\gamma+dH_{0}\right)  }+\frac{dH_{0}^{2}}{n\gamma+dH_{0}%
}\frac{F_{n}\left(  h_{i<n},\phi_{i<n},t_{i<n}\right)  -G_{n}\left(
h_{i<n},\phi_{i<n},t_{i<n}\right)  }{2n^{2}\gamma^{2}\left[  n\gamma+\left(
d-1\right)  H_{0}\right]  },\nonumber\\
h_{n}  &  =\frac{F_{n}\left(  h_{i<n},\phi_{i<n},t_{i<n}\right)  -G_{n}\left(
h_{i<n},\phi_{i<n},t_{i<n}\right)  }{2dn^{2}\gamma^{2}\beta_{1}\left[
n\gamma+\left(  d-1\right)  H_{0}\right]  },\label{eq:rere}\\
t_{n}  &  =\frac{1}{\gamma}\frac{H_{n}\left(  h_{i<n},\phi_{i<n}%
,t_{i<n}\right)  }{\left(  n^{2}-1\right)  \gamma+dH_{0}\left(  n-1\right)
}.\nonumber
\end{align}
From eqns. $\left(  \ref{eq:T-phi ansatz}\right)  $, one obtains that
$t_{1}=1$ and $\phi_{1}=0$. Since\ $\dot{H}$, $F\left(  R\right)
\sim\mathcal{O}\left(  e^{\gamma t}\right)  $ and $\dot{T}^{2},\Phi
\sim\mathcal{O}\left(  e^{2\gamma t}\right)  ,$all terms of eqn. $\left(
\ref{eq:mix 1}\right)  $ except $\frac{1}{2}\ddot{F}\left(  R\right)  $ are at
$\mathcal{O}\left(  e^{2\gamma t}\right)  $. Given $\ddot{F}\left(  R\right)
=2d\beta_{1}h_{1}\gamma^{3}e^{\gamma t}+\mathcal{O}\left(  e^{2\gamma
t}\right)  $, one has $h_{1}=0$ since $\beta_{1}\neq0$ by assumption. After
the initial conditions $h_{1}=0$, $\phi_{1}=0$, and $t_{1}=1$ are obtained,
one could use the recurrence relations $\left(  \ref{eq:rere}\right)  $ to
find values of $h_{n},\phi_{n},$and $t_{n}$. For example, we have for $n=2$
that
\begin{equation}
\phi_{2}=\frac{1}{8},h_{2}=\frac{\gamma}{4d\beta_{1}\left[  \left(
d+1\right)  H_{0}^{2}-2dH_{0}\gamma-4\gamma^{2}\right]  },t_{2}=0,
\end{equation}
where $h_{2}$ is generally not zero.

\subsection{Pole at $R_{0}$}

We now study the scenario\ in which $f\left(  R\right)  $ has a pole of order
$\left\vert L\right\vert $ at $R_{0}$. Therefore, the Laurent series expansion
of $f\left(  R\right)  $ at $R_{0}$ is%
\begin{equation}
f\left(  R\right)  =\sum\limits_{l\geq L}\alpha_{l}\left(  R-R_{0}\right)
^{l}, \label{eq:R series}%
\end{equation}
where $\alpha_{L}\neq0$ and $L\leq-1$ is some negative integer. We now
consider two cases: $R_{0}=0$ and $R_{0}\neq0$.

\subsubsection{$R_{0}=0$}

Since $H_{0}=0$, the ansatz for $H\left(  t\right)  $ $\left(
\ref{eq:H ansatz}\right)  $ becomes
\begin{equation}
H\left(  t\right)  =\underset{n\geq N}{\sum}h_{n}e^{nmt},
\label{eq:H ansatz pole}%
\end{equation}
where $h_{N}e^{Nmt}$ is assumed to be the nonzero leading term. If $H\left(
t\right)  \neq0$, one has some integer $N\geq1$ for which $h_{N}\neq0$.
Plugging eqn. $\left(  \ref{eq:H ansatz pole}\right)  $ into $R=2d\dot
{H}+d\left(  d+1\right)  H^{2}$ gives for nonzero integer $l$ that
\begin{equation}
R^{l}=r_{N}^{l}e^{Nlmt}+\mathcal{O}\left(  e^{\left(  Nl+1\right)  mt}\right)
,
\end{equation}
where $r_{N}=2dNmh_{N}$. Thus, we have%
\begin{equation}
F\left(  R\right)  =\sum\limits_{l\geq L}l\alpha_{l}R^{l-1}=Lr_{N}^{L-1}%
\alpha_{L}e^{N\left(  L-1\right)  mt}+\mathcal{O}\left(  e^{\left[  N\left(
L-1\right)  +1\right]  mt}\right)  . \label{eq:FR}%
\end{equation}
Given eqn. $(\ref{eq:FR}),$ the $ij$ components of gravitational equation
$(\ref{eq:ij component})$ becomes
\[
-Lr_{N}^{l-1}\alpha_{L}N^{2}\left(  L-1\right)  ^{2}m^{2}e^{N\left(
L-1\right)  mt}+\mathcal{O}\left(  e^{\left[  N\left(  L-1\right)  +1\right]
mt}\right)  =0,
\]
where gives $h_{N}^{L-1}\alpha_{L}=0$. Since $\alpha_{L}\neq0$, one obtains
$h_{N}=0$ and hence a contradiction, which means $H\left(  t\right)  =0$.
However $f\left(  R\right)  $ blows up at $R=0$, and hence the tachyon rolling
ansatzes $\left(  \ref{eq:T-phi ansatz}\right)  $ and $\left(
\ref{eq:H ansatz}\right)  $ do not solve the equations of motion $\left(
\ref{eq: 00 component}\right)  $, $\left(  \ref{eq:ij component}\right)  $ and
$(\ref{eq:tachyoneom})$ in this case.

\subsubsection{$R_{0}\neq0$}

The ansatz for $H\left(  t\right)  $ is
\begin{equation}
H\left(  t\right)  =H_{0}+\underset{n\geq N}{\sum}h_{n}e^{n\gamma t},
\label{eq:H ansatz pole 2}%
\end{equation}
where $h_{N}e^{N\gamma t}$ is the first nonzero term in the series. Using eqn.
$\left(  \ref{eq:R series}\right)  $, one has%
\begin{equation}
f\left(  R\right)  =\alpha_{L}r_{N}^{L}e^{NL\gamma t}+\mathcal{O}\left(
e^{\left(  NL+1\right)  \gamma t}\right)  ,
\end{equation}
where $r_{N}=2d\left[  N\gamma+\left(  d+1\right)  H_{0}\right]  h_{N}.$
Therefore, eqn. $\left(  \ref{eq: 00 component}\right)  $ gives%
\begin{equation}
\left[  H_{0}-N\left(  L-1\right)  \gamma\right]  dH_{0}Lr_{N}^{L-1}\alpha
_{L}e^{N\left(  L-1\right)  \gamma t}+\mathcal{O}\left(  e^{\left(  N\left(
L-1\right)  +1\right)  \gamma t}\right)  =0. \label{eq:ij pole}%
\end{equation}
If $N\gamma+\left(  d+1\right)  H_{0}\neq0$ and $H_{0}-N\left(  L-1\right)
\gamma\neq0$ for any positive integer $N$, eqn. $\left(  \ref{eq:ij pole}%
\right)  $ gives $\alpha_{L}=0$, which means $H\left(  t\right)  =H_{0}$.
However, $f\left(  R\right)  $ blows up at $R=R_{0}$, and hence the tachyon
rolling ansatz does not solve the equations of motion. If $N\gamma+\left(
d+1\right)  H_{0}=0$ or $H_{0}-N\left(  L-1\right)  \gamma=0$ some $N$, the
coefficient of $e^{N\left(  L-1\right)  \gamma t}$ in eqn. $\left(
\ref{eq:ij pole}\right)  $ is zero without making $\alpha_{L}=0$. In
principle, one could put the ansatz into the equations of motion and solve for
$h_{n}$, $\phi_{n}$, and $t_{n}$. However, one could not obtain the recurrence
relations as in the $F\left(  R_{0}\right)  =0$ case. One might need other
means to find values of $h_{n},\phi_{n},$and $t_{n}$.

\section{Cosmological Solutions}

\label{Sec:CS}

In section \ref{Sec:TRS}, we considered a special class of solutions to the
equations of motion, whose ansatzes are given in eqns. $\left(
\ref{eq:T-phi ansatz}\right)  $ and $\left(  \ref{eq:H ansatz}\right)  $. For
the case with $f\left(  R\right)  =R$, the string metric remains fixed for
such solutions. However for some more general form of $f\left(  R\right)  $,
there might exist tachyon-driven rolling solutions with nonzero $H\left(
t\right)  $. It appears that the behavior of the classical solutions is richer
in the $f\left(  R\right)  $ theory. In this section, we study the classical
solutions beyond the ansatzes $\left(  \ref{eq:T-phi ansatz}\right)  $ and
$\left(  \ref{eq:H ansatz}\right)  $. We are still interested in the form of
solutions in eqns. $\left(  \ref{eq:ansatz}\right)  $, in which the metric is
the spatially flat FRW metric. The equations of motion then reduce to eqns.
$(\ref{eq: 00 component})$, $(\ref{eq:ij component})$, $(\ref{eq: dilatoneom}%
)$, and $(\ref{eq:tachyoneom})$. Note that if $V\left(  T\right)  =V\left(
-T\right)  $, the solutions are invariant under the "time-reversal"
transformations $t\rightarrow-t$, for which%
\begin{equation}
H\rightarrow-H\text{, }R\rightarrow R\text{, }\Phi\rightarrow\Phi\text{,
}T\rightarrow-T.
\end{equation}
However, the equations of motion might become nonlinear higher order
differential equations, which are difficult to solve. To investigate the
properties of their solutions, we consider two simple scenarios, the one with
constant $H$, and the other with constant $T$.

\subsection{Constant $H$}

If we assume $H\left(  t\right)  =H_{0}$ which is a constant, there are three
independent equations of motion for $\Phi\left(  t\right)  $ and $T\left(
t\right)  $. Therefore, one of these equations determines the form of the
tachyon potential $V\left(  T\right)  $, which depends on $H_{0}$. In other
words, only some particular potentials $V\left(  T\right)  $ admit the
classical solutions with $H\left(  t\right)  =H_{0}$. In this case, the $ij$
components of gravitational equation $(\ref{eq:ij component})$ becomes%
\begin{equation}
dF\left(  R_{0}\right)  H_{0}^{2}-4\left[  F\left(  R_{0}\right)  -1\right]
\dot{\Phi}^{2}+2\left[  F\left(  R_{0}\right)  -1\right]  \ddot{\Phi}%
+2dH_{0}\left[  F\left(  R_{0}\right)  -1\right]  \dot{\Phi}-2H_{0}F\left(
R_{0}\right)  \dot{\Phi}=0, \label{ij eqn}%
\end{equation}
where $R\left(  t\right)  =R_{0}=d\left(  d+1\right)  H_{0}^{2}$. To solve
this equation, we can introduce the variable $\tau=dH_{0}t$. In what follows,
the dot denotes $t$-derivative, and the prime denotes $\tau$-derivative.

If $F\left(  R_{0}\right)  =1$ and $H_{0}=0$, eqn. $\left(  \ref{ij eqn}%
\right)  $ becomes trivial and hence there are only two independent equations
of motion for $\Phi\left(  t\right)  $ and $T\left(  t\right)  $. In this
case, $\Phi\left(  t\right)  $ and $T\left(  t\right)  $ can be solved for any
$V\left(  T\right)  $, and the properties of these solutions have been
discussed in \cite{Yang:2005rw}. It has been found that the string coupling
always became divergent at some time.

If $F\left(  R_{0}\right)  =1$ and $H_{0}\neq0$, the solution to eqn. $\left(
\ref{ij eqn}\right)  $ is%
\begin{equation}
\Phi\left(  \tau\right)  =\frac{\tau-\tau_{0}}{2}, \label{eq:phi(t)0}%
\end{equation}
where the integration constant $\tau_{0}$ is a constant time translation and
can be set to zero for simplicity from now on. The dilaton $\Phi$ goes to
infinity at $t=+\infty$. This solution evolves to a singular configuration
with a strongly coupled background at infinite string time. The dilaton
equation $\left(  \ref{eq: dilatoneom}\right)  $ gives $V\left(  T\right)
=\frac{1}{2}\left[  f\left(  R_{0}\right)  -R_{0}\right]  $. However, eqn.
$\left(  \ref{eq: 00 component}\right)  $ becomes%
\begin{equation}
\dot{T}^{2}=-dH_{0}^{2}<0,
\end{equation}
which contradicts the tachyon field $T$ being real. Thus, there are no
classical solutions with constant $H$ in this case.

If $F\left(  R_{0}\right)  \neq1$, solving eqn. $\left(  \ref{ij eqn}\right)
$ for $\Phi$ gives
\begin{equation}
\Phi_{\pm}\left(  \tau\right)  =\frac{1}{2}\ln\left\vert \frac{e^{\tau}}%
{1\pm\gamma e^{A\tau}}\right\vert +C_{\Phi}, \label{eq:phi(t)}%
\end{equation}
where $C_{\Phi}$ is an integration constant, $A=1+\frac{F\left(  R_{0}\right)
}{d\left[  F\left(  R_{0}\right)  -1\right]  }$, and $\gamma=\left\vert
1-F\left(  R_{0}\right)  \right\vert $. Plugging $\Phi\left(  t\right)  $ into
the dilaton equation $\left(  \ref{eq: dilatoneom}\right)  $, one could solve
it for $V\left(  T\left(  \tau\right)  \right)  $:%
\begin{equation}
\frac{V\left(  \tau\right)  }{d^{2}H_{0}^{2}}=-\frac{F\left(  R_{0}\right)
R_{0}-f\left(  R_{0}\right)  }{2d^{2}H_{0}^{2}}+sgn\left(  F\left(
R_{0}\right)  -1\right)  \frac{\rho Ae^{^{A\tau}}}{\pm1+\gamma e^{^{A\tau}}},
\label{eq:V(t)}%
\end{equation}
where $\rho=\frac{F\left(  R_{0}\right)  \left[  dF\left(  R_{0}\right)
-d+1\right]  }{2d}$, and $sgn\left(  x\right)  $ is the sign function with
$sgn\left(  x\right)  =\frac{x}{\left\vert x\right\vert }$. Plugging
$\Phi\left(  t\right)  $ into the gravitational equation
$(\ref{eq: 00 component})$ gives%
\begin{equation}
T_{\pm}^{\prime2}=\alpha\frac{\pm1+\beta e^{A\tau}}{\left(  1\pm\gamma
e^{A\tau}\right)  ^{2},} \label{eq:TDS}%
\end{equation}
where $B=d+F\left(  R_{0}\right)  -3dF\left(  R_{0}\right)  +2dF^{2}\left(
R_{0}\right)  $, $\alpha=\frac{F\left(  R_{0}\right)  \left(  d+1\right)
-d}{d}$, and $\beta=\frac{B}{2d\left\vert 1-F\left(  R_{0}\right)  \right\vert
}$. When $F\left(  R_{0}\right)  =\frac{d}{d+1}$, one has that $\alpha=A=0$.
In this case, $T$ stays constant, and
\begin{equation}
\Phi\left(  \tau\right)  =\frac{\tau}{2}+C_{\Phi},
\end{equation}
which is the linear dilaton solution. When $F\left(  R_{0}\right)  \neq
\frac{d}{d+1}$, we list the sign of $T_{\pm}^{\prime2}$ for the possible
values of $F\left(  R_{0}\right)  $ in TABLE \ref{tab:one} for $d<6$ and TABLE
\ref{tab:two}\ for $d\geq6$, where we define $F_{\pm}=\frac{-1+3d\pm
\sqrt{1-6d+d^{2}}}{4d}$ such that $B=0$ when $F\left(  R_{0}\right)  =F_{\pm}%
$. Note that $0<F_{-}<F_{+}<\frac{d}{d+1}$. When $T_{\pm}^{\prime2}\geq0$, we
integrate eqn. $\left(  \ref{eq:TDS}\right)  $ and obtain
\begin{equation}
T_{\pm}\left(  \tau\right)  =\frac{2\sqrt{\alpha}}{A}\operatorname{Re}\left[
\sqrt{\frac{\gamma-\beta}{\gamma}}\text{arctanh}\left(  \sqrt{\frac{\gamma
}{\gamma-\beta}}\sqrt{\pm1+\beta e^{A\tau}}\right)  -\text{arctanh}\left(
\sqrt{\pm1+\beta e^{A\tau}}\right)  \right]  +C_{T}, \label{eq:T(t)}%
\end{equation}
where $C_{T}$ is an integration constant. In principle, one could use eqns.
$\left(  \ref{eq:V(t)}\right)  $ and $\left(  \ref{eq:T(t)}\right)  $ to find
$V\left(  T\right)  $.

\begin{table}[tb]
\begin{center}
$%
\begin{tabular}
[c]{|c|c|c|}\hline
& $F\left(  R_{0}\right)  >\frac{d}{d+1}$ & $F\left(  R_{0}\right)  <\frac
{d}{d+1}$\\\hline
$T_{+}^{\prime2}$ & $T_{+}^{\prime2}>0$ & $T_{+}^{\prime2}<0$\\\hline
$T_{-}^{\prime2}$ & $T_{-}^{\prime2}\geq\left(  <\right)  0$ for $e^{A\tau
}\geq\left(  <\right)  \frac{2d\gamma}{B}$ & $T_{-}^{\prime2}\geq\left(
<\right)  0$ for $e^{A\tau}\leq\left(  >\right)  \frac{2d\gamma}{B}$\\\hline
\end{tabular}
\ \ \ \ $
\end{center}
\caption{The sign of $T_{\pm}^{\prime2}$ for $d<6$.}%
\label{tab:one}%
\end{table}

\begin{table}[tb]
\begin{center}
$%
\begin{tabular}
[c]{|c|c|c|c|}\hline
& $F\left(  R_{0}\right)  >\frac{d}{d+1}$ & $F_{+}\leq F\left(  R_{0}\right)
<\frac{d}{d+1}$ or $F\left(  R_{0}\right)  \leq F_{-}$ & $F_{-}<F\left(
R_{0}\right)  <F_{+}$\\\hline
$T_{+}^{\prime2}$ & $T_{+}^{\prime2}>0$ & $T_{+}^{\prime2}<0$ & $T_{+}%
^{\prime2}\geq\left(  <\right)  0$ for $e^{A\tau}\geq\left(  <\right)
\frac{2d\gamma}{\left\vert B\right\vert }$\\\hline
$T_{-}^{\prime2}$ & $T_{-}^{\prime2}\geq\left(  <\right)  0$ for $e^{A\tau
}\geq\left(  <\right)  \frac{2d\gamma}{\left\vert B\right\vert }$ &
$T_{-}^{\prime2}\geq\left(  <\right)  0$ for $e^{A\tau}\leq\left(  >\right)
\frac{2d\gamma}{\left\vert B\right\vert }$ & $T_{-}^{\prime2}<0$\\\hline
\end{tabular}
\ \ \ \ $
\end{center}
\caption{The sign of $T_{\pm}^{\prime2}$ for $d\geq6$.}%
\label{tab:two}%
\end{table}

\begin{figure}[tb]
\begin{center}
\subfigure[{~Plot of $\Phi_{+}\left(  \tau\right)  $}]
{\includegraphics[width=0.38\textwidth]{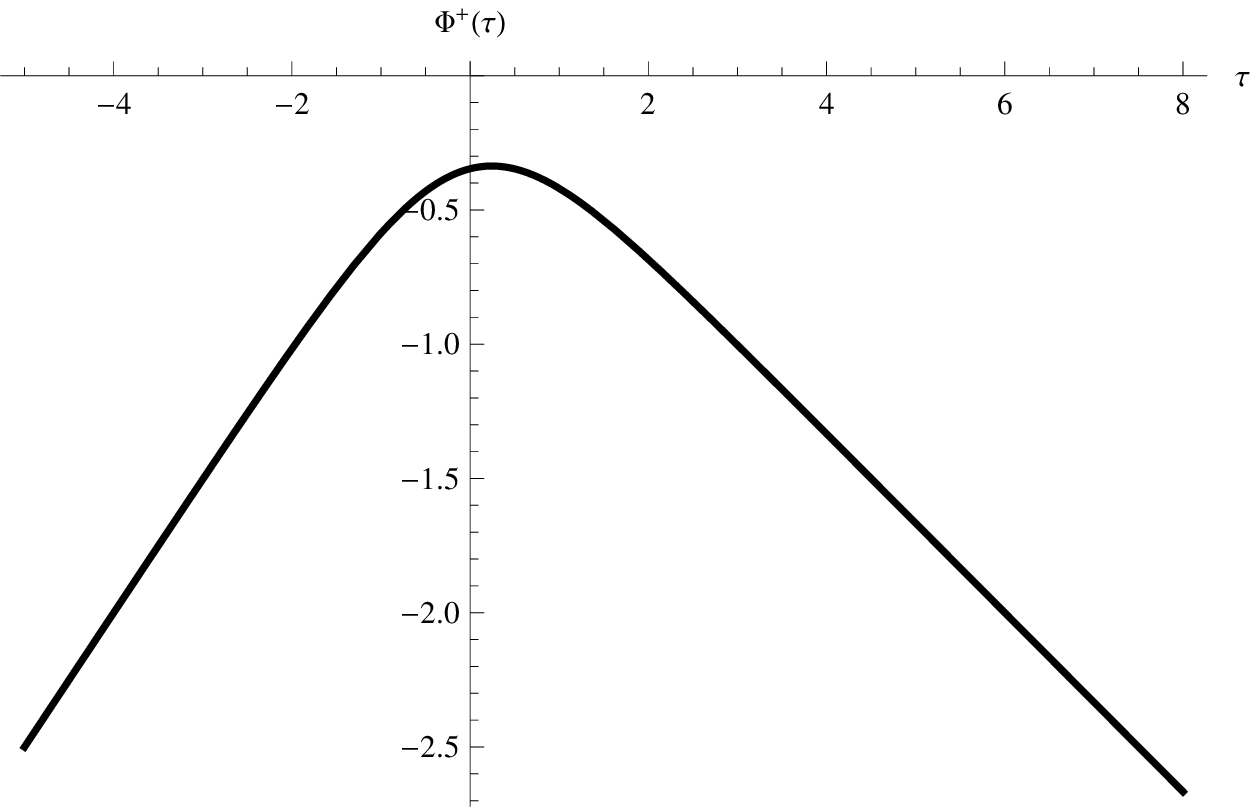}\label{fig:PhiVPlus:a}}
\subfigure[{~Plot of $V\left(  T\right)  $}]
{\includegraphics[width=0.52\textwidth]{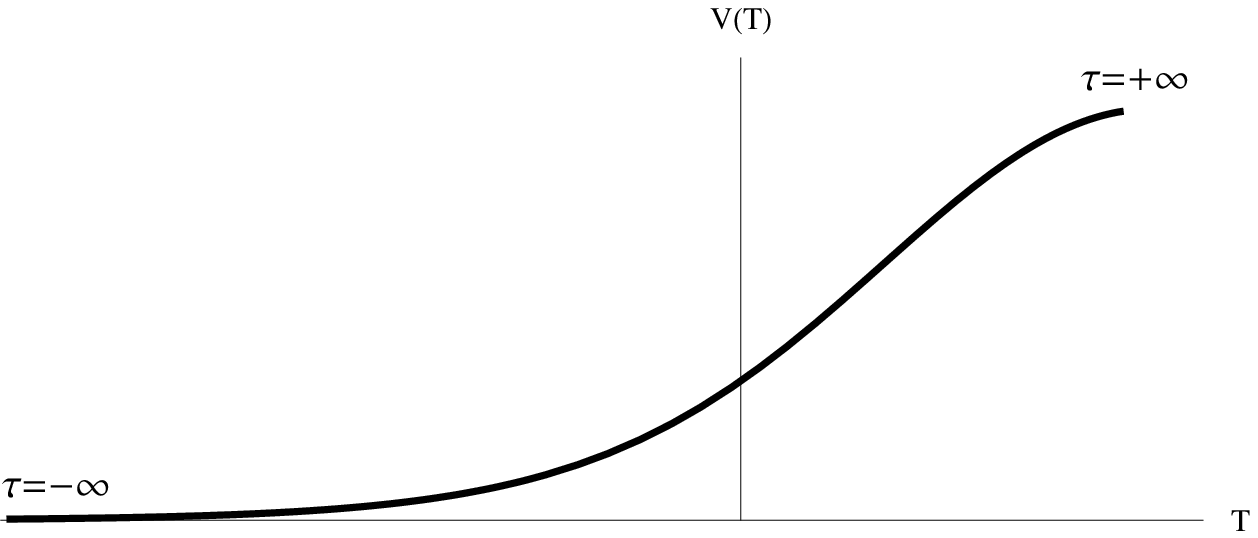}\label{fig:PhiVPlus:b}}
\end{center}
\caption{Plots of $\Phi_{+}\left(  \tau\right)  $ and $V\left(  T\right)  $
for $F\left(  R_{0}\right)  =2$, where we have $C_{\Phi}=0$.}%
\label{fig:PhiVPlus}%
\end{figure}

When $\tau\rightarrow-\infty$, $\Phi_{\pm}\left(  \tau\right)  $ always goes
to $-\infty$. Defining $\tau=\tau_{a}$ such that $\gamma e^{^{A\tau_{a}}}=1$,
one finds that $\Phi_{-}\left(  \tau_{a}\right)  =+\infty$. As $\tau
\rightarrow+\infty$, one has for $\Phi_{\pm}\left(  \tau\right)  $ that%
\begin{equation}
\Phi_{\pm}\left(  \tau\right)  \rightarrow\left\{
\begin{array}
[c]{c}%
+\infty\text{, \ \ \ \ \ for }A<1\\
\text{ }C_{\Phi}\text{, \ \ \ \ \ for }A=1\\
-\infty\text{, \ \ \ \ \ for }A>1
\end{array}
\text{.}\right.
\end{equation}
For $\Phi_{+}\left(  \tau\right)  \ $with $A<1$ and $\Phi_{-}\left(
\tau\right)  $, the string coupling always diverges at some time. However for
$\Phi_{+}\left(  \tau\right)  \ $with $A\geq1$, the string coupling always
stay finite. In FIG. $\ref{fig:PhiVPlus:a}$, we plot $\Phi_{+}\left(
\tau\right)  $ for $F\left(  R_{0}\right)  =2$. Note that $A\geq1$ implies
that $F\left(  R_{0}\right)  \leq0$ or $F\left(  R_{0}\right)  >1$. TABLEs
\ref{tab:one} and \ref{tab:two} show that $T_{+}^{\prime2}<0$ when $F\left(
R_{0}\right)  \leq0$. Thus when $A\geq1$, the solution $T_{+}\left(
\tau\right)  $ only exists for $F\left(  R_{0}\right)  >1$. When $F\left(
R_{0}\right)  >1$, eqns. $\left(  \ref{eq:V(t)}\right)  $ and $\left(
\ref{eq:T(t)}\right)  $ shows that as $\tau\rightarrow-\infty$%
\begin{equation}
T_{+}\left(  \tau\right)  \sim\sqrt{\alpha}\tau\text{, }\frac{dV\left(
T\right)  }{dT}\sim d^{2}H_{0}^{2}\frac{\rho A^{2}}{\sqrt{\alpha}}e^{^{A\tau}%
}\text{, and }\frac{d^{2}V\left(  T\right)  }{dT^{2}}\sim d^{2}H_{0}^{2}%
\frac{\rho A^{4}}{\alpha}e^{A\tau},\nonumber
\end{equation}
and as $\tau\rightarrow+\infty$%
\begin{equation}
T_{+}\left(  \tau\right)  \sim C_{T}\text{, }\frac{dV\left(  T\right)  }%
{dT}\sim d^{2}H_{0}^{2}\frac{\rho A^{2}}{\sqrt{\alpha\beta}\gamma}e^{-A\tau
/2}\sim0\text{, and}\frac{d^{2}V\left(  T\right)  }{dT^{2}}\sim-d^{2}H_{0}%
^{2}\frac{\rho A^{4}}{2\alpha\beta}<0,\nonumber
\end{equation}
where we use $\alpha>0$, $\beta>0$, and $\rho>0$ for $F\left(  R_{0}\right)
>1$. Therefore when $\tau=+\infty$, the tachyon field $T$ stays at a maximum
of $V\left(  T\right)  $. Since $T_{+}\left(  \tau\right)  $ goes to $-\infty$
as $\tau\rightarrow-\infty$, one finds for $T\rightarrow-\infty$ that the
tachyon potential becomes
\begin{equation}
V\left(  T\right)  \sim-\frac{F\left(  R_{0}\right)  R_{0}-f\left(
R_{0}\right)  }{2}+\rho Ad^{2}H_{0}^{2}e^{^{AT/\sqrt{\alpha}}}.
\label{eq: VTinfM}%
\end{equation}
In FIG. $\ref{fig:PhiVPlus:b}$, we plot $V\left(  T\right)  $ for $F\left(
R_{0}\right)  =2$. Recalling that $\tau=dH_{0}t$, we find for $H_{0}<0$ and
$F\left(  R_{0}\right)  >1$, that the solution $T_{+}\left(  t\right)  $ in
eqn. $\left(  \ref{eq:T(t)}\right)  $ describes the scenario in which the
tachyon field is at a maximum of $V\left(  T\right)  $ at $t=-\infty$ and
begins to roll down $V\left(  T\right)  $ afterwards. At $t=\infty$,
$T_{+}\left(  t\right)  $ goes to $-\infty$, and $V\left(  T\right)  $ is
given in eqn. $\left(  \ref{eq: VTinfM}\right)  $. Moreover, this tachyon
rolling down scenario is free of singularities since the Ricci scalar is
constant, and the string coupling is always finite.

\subsection{Constant $T$}

If $T=T_{0}$ such that $V^{\prime}\left(  T_{0}\right)  =0$, the tachyon
equation $\left(  \ref{eq:tachyoneom}\right)  $ becomes trivial. As a result,
one could solve the gravitational equations $(\ref{eq: 00 component})$ and
$(\ref{eq:ij component})$ for $H\left(  t\right)  $ and $\Phi\left(  t\right)
$. The dilaton equation $\left(  \ref{eq: dilatoneom}\right)  $ impose
constraints on the integration constants. This scenario could provide some
insights into the possible final state of bulk tachyon condensation. The case
with $f\left(  R\right)  =R$ has been discussed in \cite{Suyama:2008hk}, where
the classical solutions always were found to evolve from or to singular
configurations. We here investigate the singular behavior of the solutions in
the case with $f\left(  R\right)  =R+\frac{\alpha}{2}R^{2}$, in which the
perturbation method is used to find solutions. The perturbation method gives
how these solutions are altered for non-zero but small $\alpha$. In doing so,
we assume that the altered solutions can be Taylor expanded in $\alpha$. In
addition, it turns out that the forms of the solutions depend on the sign of
$V\left(  T_{0}\right)  $. We calculate the perturbative solutions to
$\mathcal{O}\left(  \alpha\right)  $ for $V\left(  T_{0}\right)  =0$ and
discuss the singular behavior of the solutions for\ $V\left(  T_{0}\right)
>0$ and $V\left(  T_{0}\right)  <0$.

Substituting the Taylor expansions of $H\left(  t\right)  $ and $\Phi\left(
t\right)  $ in powers of $\alpha$%
\begin{align}
H\left(  t\right)   &  =H_{0}\left(  t\right)  +\alpha H_{1}\left(  t\right)
+\mathcal{O}\left(  \alpha^{2}\right)  \text{, }\nonumber\\
\Phi\left(  t\right)   &  =\Phi_{0}\left(  t\right)  +\alpha\Phi_{1}\left(
t\right)  +\mathcal{O}\left(  \alpha^{2}\right)  , \label{eq:H-Phiseries}%
\end{align}
into the gravitational equations $\left(  \ref{eq: 00 component}\right)  $ and
$\left(  \ref{eq:ij component}\right)  $, one finds
\begin{align}
d\dot{H}_{0}+dH_{0}^{2}-2\ddot{\Phi}_{0}  &  =0,\nonumber\\
\dot{H}_{0}+dH_{0}^{2}-2H_{0}\dot{\Phi}_{0}  &  =0,\nonumber\\
d\dot{H}_{1}+2dH_{0}H_{1}-2\ddot{\Phi}_{1}  &  =F\left(  t\right)
,\label{eq:00first}\\
\dot{H}_{1}+2dH_{0}H_{1}-2H_{1}\dot{\Phi}_{0}-2H_{0}\dot{\Phi}_{1}  &
=G\left(  t\right)  ,\nonumber
\end{align}
where we define%
\begin{align}
F\left(  t\right)   &  \equiv-d\left[  R_{0}\left(  \dot{H}_{0}+H_{0}%
^{2}\right)  +\left(  2R_{0}\dot{\Phi}_{0}-\dot{R}_{0}\right)  H_{0}\right]
,\nonumber\\
G\left(  t\right)   &  \equiv-R_{0}\left(  \dot{H}_{0}+dH_{0}^{2}\right)
+4R_{0}\dot{\Phi}_{0}^{2}-4\dot{\Phi}_{0}\dot{R}_{0}-2R_{0}\ddot{\Phi}%
_{0}-2\left(  d-1\right)  R_{0}H_{0}\dot{\Phi}_{0}+\ddot{R}_{0}+\left(
d-1\right)  H_{0}\dot{R}_{0}.
\end{align}

\begin{figure}[tb]
\begin{center}
\subfigure[{~Plot of $a_{+}\left(  t\right) $}]
{\includegraphics[width=0.48\textwidth]{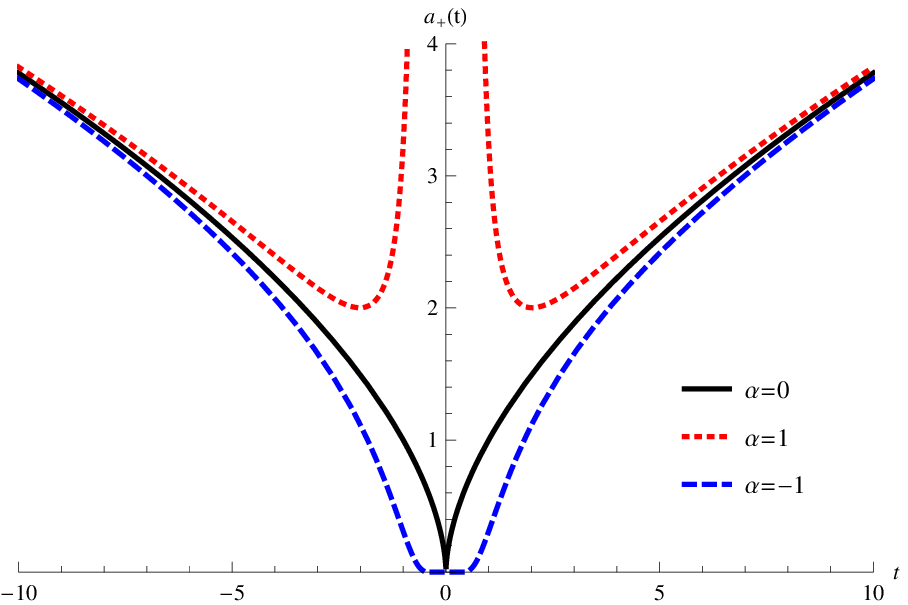}\label{fig:aPhiT:a}}
\subfigure[{~Plot of $\Phi_{+}\left(  t\right) $}]
{\includegraphics[width=0.48\textwidth]{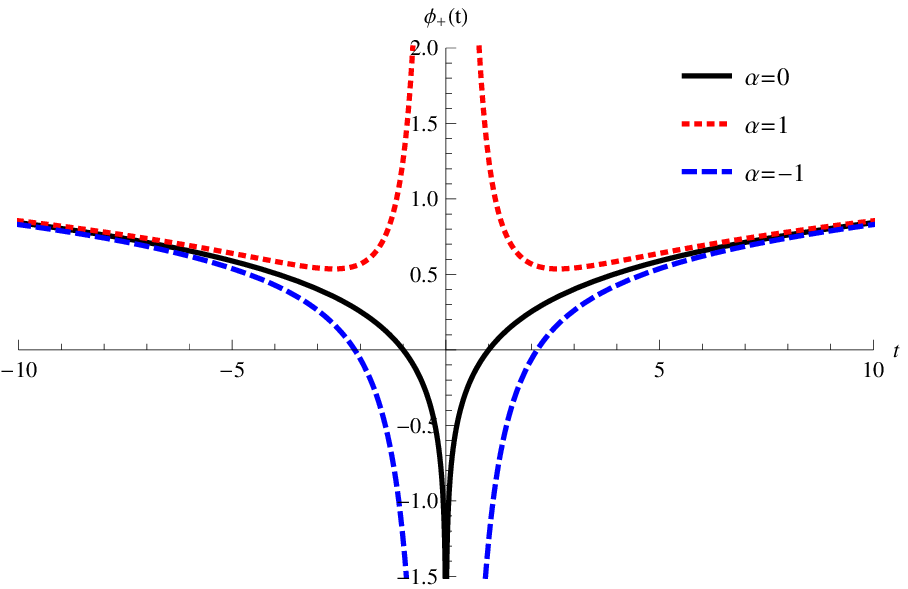}\label{fig:aPhiT:b}}
\subfigure[{~Plot of $a_{-}\left(  t\right) $}]
{\includegraphics[width=0.48\textwidth]{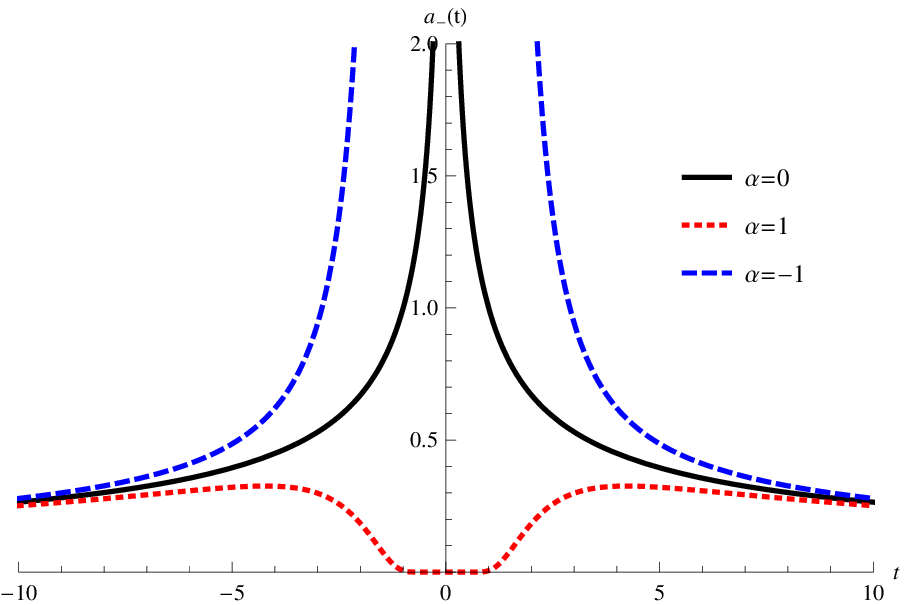}\label{fig:aPhiT:c}}
\subfigure[{~Plot of $\Phi_{-}\left(  t\right) $}]
{\includegraphics[width=0.48\textwidth]{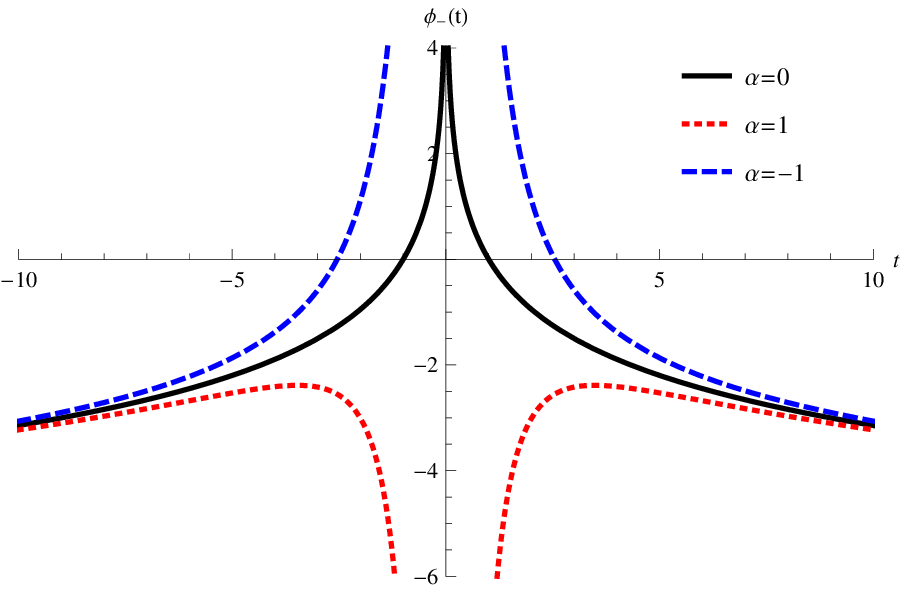}\label{fig:aPhiT:d}}
\end{center}
\caption{Plots of $a_{\pm}\left(  t\right)  $ and $\Phi_{\pm}\left(  t\right)
$, where we have $d=3$, $\alpha=1$, $t_{0}=0$, $\Phi_{0}=0$, $a_{0}=1$ and
$c=0$.}%
\label{fig:aPhiT}%
\end{figure}

If $V\left(  T_{0}\right)  =0$, solving eqns. $\left(  \ref{eq:00first}%
\right)  $ and using eqn. $\left(  \ref{eq: dilatoneom}\right)  $ to constrain
the integration constants, one finds that the solutions to $\mathcal{O}\left(
\alpha\right)  $ are%
\begin{align}
H_{\pm}\left(  t\right)   &  =\pm\frac{1}{\sqrt{d}\left(  t-t_{0}\right)
}+\frac{\alpha h^{\pm}}{4\left(  t-t_{0}\right)  ^{3}}+\alpha c\left(
t-t_{0}\right)  ^{-2},\nonumber\\
\Phi_{\pm}\left(  t\right)   &  =\frac{\pm\sqrt{d}-1}{2}\ln\left\vert
t-t_{0}\right\vert +\Phi_{0}+\frac{\alpha\phi^{\pm}}{\left(  t-t_{0}\right)
^{2}}-\frac{\alpha c\left(  d\mp\sqrt{d}\right)  }{2\left(  t-t_{0}\right)  },
\end{align}
where $c$ and $\Phi_{0}$ and are integration constants, and we have%
\begin{align}
h^{\pm}  &  =\mp\frac{\left(  \mp\sqrt{d}+1\right)  ^{2}\left(  5\pm6\sqrt
{d}+5d\right)  }{\sqrt{d}},\nonumber\\
\phi^{\pm}  &  =\frac{\left(  \pm\sqrt{d}-1\right)  ^{2}\left(  \pm
5d^{\frac{3}{2}}+4d\pm\sqrt{d}-2\right)  }{16}.
\end{align}
Integrating $H_{\pm}\left(  t\right)  $, we find%
\begin{equation}
a_{\pm}\left(  t\right)  =a_{0}\left\vert t-t_{0}\right\vert ^{\frac{\pm
1}{\sqrt{d}}}\exp\left[  -\frac{\alpha h^{\pm}}{8\left(  t-t_{0}\right)  ^{2}%
}-\frac{\alpha c}{t-t_{0}}\right]  ,
\end{equation}
where $a_{0}$ is a constant. The $\mathcal{O}\left(  \alpha\right)  $
corrections would not change the asymptotic behavior of the solutions at
$t=\pm\infty$. Around $t=t_{0}$, the $\mathcal{O}\left(  \alpha\right)  $
corrections could dramatically change the asymptotic behavior of the these
solutions. We plot $a_{\pm}\left(  t\right)  $ and $\Phi_{\pm}\left(
t\right)  $ in FIG. $\ref{fig:aPhiT}$, where we have $d=3$, $\alpha=1$,
$t_{0}=0$, $\Phi_{0}=0$, $a_{0}=1$ and $c=0$. For example, the solutions
$a_{+}\left(  t\right)  $ and $\Phi_{+}\left(  t\right)  $ with $\alpha=0$
evolve from (to) big bang (big crunch) at $t=t_{0}$, while they have a very
weakly string coupled background. However for $\alpha>0$, $a_{+}\left(
t\right)  $ goes to infinity at $t=t_{0}$, while the string coupling becomes
divergent. Note that these perturbative solution are valid when $\alpha\left(
t-t_{0}\right)  ^{-2}\ll1$. Therefore, the higher order corrections are
necessary to study the singular behavior of classical solutions. However, our
analysis gives a sense of how quantum effects modify the singular behavior of
classical solutions.

For $V\left(  T_{0}\right)  >0$, the leading terms of the solutions are
\begin{align}
H_{0}^{\pm}\left(  t\right)   &  =\pm\frac{C}{\sqrt{d}}\frac{1}{\sinh C\left(
t-t_{0}\right)  },\nonumber\\
\Phi_{0}^{\pm}\left(  t\right)   &  =\frac{1}{2}\left(  \pm\sqrt{d}-1\right)
\ln\left\vert \sinh\frac{C}{2}\left(  t-t_{0}\right)  \right\vert -\frac{1}%
{2}\left(  \pm\sqrt{d}+1\right)  \ln\left\vert \cosh\frac{C}{2}\left(
t-t_{0}\right)  \right\vert +\Phi_{0},
\end{align}
where $C=\sqrt{2V\left(  T_{0}\right)  }$, and $t_{0}$ and $\Phi_{0}$ are
integration constants. For $V\left(  T_{0}\right)  <0$, the leading terms are%
\begin{align}
H_{0}^{\pm}\left(  t\right)   &  =\pm\frac{C}{\sqrt{d}}\frac{1}{\sin C\left(
t-t_{0}\right)  },\nonumber\\
\Phi_{0}^{\pm}\left(  t\right)   &  =\frac{1}{2}\left(  \pm\sqrt{d}-1\right)
\ln\left\vert \sin\frac{C}{2}\left(  t-t_{0}\right)  \right\vert -\frac{1}%
{2}\left(  \pm\sqrt{d}+1\right)  \ln\left\vert \cos\frac{C}{2}\left(
t-t_{0}\right)  \right\vert +\Phi_{0},
\end{align}
where $C=\sqrt{-2V\left(  T_{0}\right)  }$, and $t_{0}$ and $\Phi_{0}$ are
integration constants. These solutions have a singularity at $t=t_{0}$. Around
$t=t_{0}$, their singular behaviors are the same as in the case with $V\left(
T_{0}\right)  =0$. Therefore, the last two equations in eqns. $\left(
\ref{eq:00first}\right)  $ give that the singular behaviors of $H_{1}\left(
t\right)  $ and $\Phi_{1}\left(  t\right)  $ at $t=t_{0}$ are also the same as
in the case with $V\left(  T_{0}\right)  =0$.

\section{Conclusion}

\label{Sec:CON}

In \cite{Yang:2005rw}, the tachyon-induced rolling solutions have been
considered using the low-energy effective field equations, which were derived
from the effective action $\left(  \ref{eq:action}\right)  $ with $f\left(
R\right)  =R$. To gain some insight into quantum effects on the tachyon
dynamics, in this paper we investigated the behavior of the classical
solutions of the low-energy effective action $\left(  \ref{eq:action}\right)
$\ of the graviton-dilaton-tachyon system, in which quantum corrections are
included only in $f\left(  R\right)  $ for simplicity. After the equations of
motion were obtained in section \ref{Sec:CSF}, we solved them for the
tachyon-induced rolling ansatzes $\left(  \ref{eq:T-phi ansatz}\right)  $ and
$\left(  \ref{eq:H ansatz}\right)  $ in section \ref{Sec:TRS}. Finally, more
classical solutions were discussed in section \ref{Sec:CS}.

In \cite{Yang:2005rw} where $f\left(  R\right)  =R$, it showed that $H\left(
t\right)  $ vanished identically for the tachyon-induced rolling solutions.
For more general forms of $f\left(  R\right)  $, we found in subsection
\ref{subSec:R0 analytic} that there were some cases in which $H\left(
t\right)  $ could be nonzero. Moreover, we solved the equations of motion
assuming $H\left(  t\right)  =H_{0}$ which was a constant. When $F\left(
R_{0}\right)  =1$ and $H_{0}=0$, the properties of classical solutions have
been discussed in \cite{Yang:2005rw}, and it has been found that the dilaton
always rolled toward stronger coupling. However for $F\left(  R_{0}\right)
>1$, we found that there existed some solutions in which the string coupling
could always stay finite. In the case of $f\left(  R\right)  =R+\frac{\alpha
}{2}R^{2}$, we also solved the equations of motion assuming $T\left(
t\right)  =T_{0}$, whose scenario is related to the possible final state of
bulk tachyon condensation. It turned out for some solutions that higher order
terms in $f\left(  R\right)  $ could dramatically change their singular
behavior. Since we only used an effective model, the $f\left(  R\right)  $
gravity theory, to investigate quantum effects on tachyon dynamics, one might
not take our analysis too seriously. However, our analysis suggests that
quantum corrections should be important to understand tachyon dynamics.

\appendix

\section{Appendix}

On defining%
\begin{gather}
G_{1}\equiv F\left(  R\right)  \left(  d\dot{H}+dH^{2}\right)  +\dot{T}%
^{2}-2\ddot{\Phi}+\left[  2\left(  F\left(  R\right)  -1\right)  \dot{\Phi
}-\dot{F}\left(  R\right)  \right]  dH,\label{G1}\\
G_{2}\equiv F\left(  R\right)  \left(  \dot{H}+dH^{2}\right)  -4\left[
F\left(  R\right)  -1\right]  \dot{\Phi}^{2}+4\dot{\Phi}\dot{F}\left(
R\right)  +2\left[  F\left(  R\right)  -1\right]  \ddot{\Phi}\nonumber\\
+2\left[  F\left(  R\right)  \left(  d-1\right)  -d\right]  H\dot{\Phi}%
-\ddot{F}\left(  R\right)  -\left(  d-1\right)  H\dot{F}\left(  R\right)  ,\\
G_{3}\equiv\frac{1}{2}\left[  F\left(  R\right)  R-f\left(  R\right)  \right]
-2\left\{  d\left[  F\left(  R\right)  -1\right]  +1\right\}  \dot{\Phi}%
^{2}+2d\dot{\Phi}\dot{F}\left(  R\right) \label{G3}\\
+\left\{  d\left[  F\left(  R\right)  -1\right]  +1\right\}  \left(
\ddot{\Phi}+dH\dot{\Phi}\right)  -\frac{1}{2}d\left[  \ddot{F}\left(
R\right)  +dH\dot{F}\left(  R\right)  \right]  +V\left(  T\right)
,\nonumber\\
G_{4}\equiv\ddot{T}+\left(  dH-2\dot{\Phi}\right)  \dot{T}+V^{\prime}\left(
T\right)  , \label{G4}%
\end{gather}
the equations of motion $(\ref{eq: 00 component})$, $(\ref{eq:ij component})$,
$(\ref{eq: dilatoneom})$, and $(\ref{eq:tachyoneom})$ become $G_{i}=0$.
Supposing that $G_{1}=G_{2}=G_{3}=0$, we now show that this leads to
$G_{4}=0\,\ $whenever $\dot{T}\neq0$.

Differentiating $G_{1}=0$ with respect to time gives%
\begin{align}
\dot{T}\ddot{T}  &  =-\frac{\dot{F}\left(  R\right)  \left(  d\dot{H}%
+dH^{2}\right)  }{2}-\frac{F\left(  R\right)  \left(  d\ddot{H}+2dH\dot
{H}\right)  }{2}+\dddot{\Phi}\nonumber\\
&  -\left\{  \left[  F\left(  R\right)  -1\right]  \dot{\Phi}-\frac{\dot
{F}\left(  R\right)  }{2}\right\}  d\dot{H}-\left\{  \left[  F\left(
R\right)  -1\right]  \ddot{\Phi}+\dot{F}\left(  R\right)  \dot{\Phi}%
-\frac{\ddot{F}\left(  R\right)  }{2}\right\}  dH, \label{G2}%
\end{align}
\newline where we use $R=2d\dot{H}+d\left(  d+1\right)  H^{2}.$ By multiplying
$G_{4}$ by $\dot{T}$, eliminating $\dot{T}\ddot{T}$ through eqn. $\left(
\ref{G2}\right)  $ and $\dot{T}^{2}$ through the equation of motion $G_{1}=0$,
we find
\begin{gather}
\dot{T}G_{4}-\left(  dH-2\dot{\Phi}\right)  G_{1}-\frac{1}{2}\frac{dG_{1}}%
{dt}=-\frac{\dot{F}\left(  R\right)  \left(  d\dot{H}+dH^{2}\right)  }%
{2}-\frac{F\left(  R\right)  \left(  d\ddot{H}+2dH\dot{H}\right)  }{2}%
+\dddot{\Phi}\nonumber\\
-\left\{  \left[  F\left(  R\right)  -1\right]  \dot{\Phi}-\frac{\dot
{F}\left(  R\right)  }{2}\right\}  d\dot{H}-\left\{  \left[  F\left(
R\right)  -1\right]  \ddot{\Phi}+\dot{F}\left(  R\right)  \dot{\Phi}%
-\frac{\ddot{F}\left(  R\right)  }{2}\right\}  dH\label{G5}\\
-\left(  dH-2\dot{\Phi}\right)  \left\{  F\left(  R\right)  \left(  d\dot
{H}+dH^{2}\right)  -2\ddot{\Phi}+\left[  2\left(  F\left(  R\right)
-1\right)  \dot{\Phi}-\dot{F}\left(  R\right)  \right]  dH\right\}
+V^{\prime}\left(  T\right)  \dot{T}=0.\nonumber
\end{gather}
Differentiating $G_{3}=0$ with respect to time gives%
\begin{gather}
\frac{\dot{F}\left(  R\right)  R}{2}-4\left[  d\left(  F\left(  R\right)
-1\right)  +1\right]  \dot{\Phi}\ddot{\Phi}-2d\dot{F}\left(  R\right)
\dot{\Phi}^{2}+2d\ddot{\Phi}\dot{F}\left(  R\right)  +2d\dot{\Phi}\ddot
{F}\left(  R\right)  +d\dot{F}\left(  R\right)  \left(  \ddot{\Phi}%
+dH\dot{\Phi}\right) \nonumber\\
+\left\{  d\left[  F\left(  R\right)  -1\right]  +1\right\}  \left(
\dddot{\Phi}+d\dot{H}\dot{\Phi}+dH\ddot{\Phi}\right)  -\frac{1}{2}d\left[
\dddot{F}\left(  R\right)  +d\dot{H}\dot{F}\left(  R\right)  +dH\ddot
{F}\left(  R\right)  \right]  =-V^{\prime}\left(  T\right)  \dot{T}.
\label{G9}%
\end{gather}
Differentiating $G_{2}=0$ with respect to time and multiplying by $d/2$ on
both sides gives%
\begin{gather}
d\left[  F\left(  R\right)  -1\right]  \dddot{\Phi}-\frac{d}{2}\dddot
{F}\left(  R\right)  -4d\left[  F\left(  R\right)  -1\right]  \dot{\Phi}%
\ddot{\Phi}-2d\dot{F}\left(  R\right)  \dot{\Phi}^{2}+2d\ddot{\Phi}\dot
{F}\left(  R\right) \nonumber\\
+2d\dot{\Phi}\ddot{F}\left(  R\right)  +d\dot{F}\left(  R\right)  \ddot{\Phi
}+d^{2}H\dot{\Phi}\dot{F}\left(  R\right)  -\frac{d^{2}}{2}\dot{H}\dot
{F}\left(  R\right)  -\frac{d^{2}}{2}H\ddot{F}\left(  R\right) \label{G10}\\
=-\frac{d\dot{F}\left(  R\right)  \left(  \dot{H}+dH^{2}\right)  }{2}%
-\frac{dF\left(  R\right)  \left(  \ddot{H}+2dH\dot{H}\right)  }{2}%
+dH\dot{\Phi}\dot{F}\left(  R\right) \nonumber\\
-d\left[  F\left(  R\right)  \left(  d-1\right)  -d\right]  \dot{H}\dot{\Phi
}-d\left[  F\left(  R\right)  \left(  d-1\right)  -d\right]  H\ddot{\Phi
}-\frac{d\dot{H}\dot{F}\left(  R\right)  }{2}-\frac{dH\ddot{F}\left(
R\right)  }{2}.\nonumber
\end{gather}
Subtracting eqn. $\left(  \ref{G10}\right)  $ from eqn. $\left(
\ref{G9}\right)  $ gives
\begin{gather}
\frac{dG_{3}}{dt}-\frac{d}{2}\frac{dG_{2}}{dt}=\dddot{\Phi}-4\dot{\Phi}%
\ddot{\Phi}+\left[  F\left(  R\right)  +1\right]  d\dot{H}\dot{\Phi}%
-\frac{dF\left(  R\right)  \ddot{H}}{2}\label{G7}\\
+\left[  F\left(  R\right)  +1\right]  dH\ddot{\Phi}-\frac{dH\ddot{F}\left(
R\right)  }{2}+dH\dot{\Phi}\dot{F}\left(  R\right)  -d^{2}F\left(  R\right)
H\dot{H}+\frac{d\dot{F}\left(  R\right)  H^{2}}{2}+V^{\prime}\left(  T\right)
\dot{T}=0.\nonumber
\end{gather}
Subtracting eqn. $\left(  \ref{G5}\right)  $ from eqn. $\left(  \ref{G7}%
\right)  $ gives
\begin{gather}
\frac{dG_{3}}{dt}-\frac{d}{2}\frac{dG_{2}}{dt}-\dot{T}G_{4}+\left(
dH-2\dot{\Phi}\right)  G_{1}+\frac{1}{2}\frac{dG_{1}}{dt}\nonumber\\
=dH\left\{  2\left(  F\left(  R\right)  -1\right)  \ddot{\Phi}-\ddot{F}\left(
R\right)  +F\left(  R\right)  \left(  \dot{H}+dH^{2}\right)  -\left(
d-1\right)  H\dot{F}\left(  R\right)  \right. \label{G11}\\
\left.  -4\left[  F\left(  R\right)  -1\right]  \dot{\Phi}^{2}+4\dot{\Phi}%
\dot{F}\left(  R\right)  +2\left[  F\left(  R\right)  \left(  d-1\right)
-d\right]  H\dot{\Phi}\right\}  .\nonumber
\end{gather}
Comparing eqn. $\left(  \ref{G11}\right)  $ with eqn. $\left(  \ref{G4}%
\right)  $, we obtain the relation for $G_{1}$, $G_{2}$, $G_{3}$ and $G_{4}$%
\begin{equation}
\left(  dH-2\dot{\Phi}\right)  G_{1}+\frac{1}{2}\frac{dG_{1}}{dt}-\frac{d}%
{2}\frac{dG_{2}}{dt}-dHG_{2}+\frac{dG_{3}}{dt}-\dot{T}G_{4}=0, \label{G12}%
\end{equation}
which shows that $G_{4}=0$ if $G_{1}=G_{2}=G_{3}=0$ and $\dot{T}\neq0$.

\begin{acknowledgments}
We are grateful to Jun Tao and Zheng Sun for useful discussions. This work is
supported in part by NSFC (Grant No. 11005016, 11175039 and 11375121).
\end{acknowledgments}

\end{document}